\documentclass[aip,jcp,reprint,twocolumn]{revtex4-2}

\usepackage{amsmath,amssymb}
\usepackage{graphicx}
\usepackage{bm}
\usepackage{dcolumn}
\usepackage[colorlinks=true,linkcolor=blue,citecolor=blue,urlcolor=blue]{hyperref}
\usepackage{fancyhdr}

\begin{document}

\title{Field closure, ice neurons, and when a dendrite is a motif}

\author{Nima Dehghani}
\affiliation{McGovern Institute for Brain Research, Massachusetts Institute of Technology, Cambridge, Massachusetts, USA}
\affiliation{The NSF AI Institute for Artificial Intelligence and Fundamental Interactions (IAIFI), Cambridge, Massachusetts, USA}
\email{nima.dehghani@mit.edu}

\date{\today}

\begin{abstract}
Dendritic silhouettes are cheap. Pond ice neurons, diffusion-limited aggregates, wiring-minimizing arbors, and biological neurons can share a branching statistic because they share a growth instability, not because they are functionally similar. We take that observation as a modeling constraint. The ice neuron only structurally resembles a biological neuron. A biological neuron carries an internal state and a field that it helps generate, letting geometry, dynamics, and field co-sculpt one another. \\
In our construction, the ice neuron is the zero-gain limit of an interfacial system in which a FitzHugh--Nagumo state lives on the moving front and writes back into the field that drives it, with dimensionless gain $\gamma$. At $\gamma=0$, morphology and internal dynamics decouples into a Mullins--Sekerka growth band and a damped oscillator. Finite $\gamma$ deforms the spectrum, shifts the selected instability, and opens an oscillatory region below the isolated Hopf threshold.\\
On a dendrite grown at $\gamma=0$ and then frozen, field closure increases distal persistence without altering the geometry; its first-order effect remains positive across repeated realizations of the same growth rule and localizes to a small set of non-cable field shortcuts. Nonlinear dynamics remain finite beyond the linear instability: below threshold, closure is almost the transmission; at spike amplitude, it is a correction to a regenerative cable event.\\
A dendritic tree therefore is not intrinsically a structural spandrel. Shape alone is morphology without an internal state or a source term. That is the dendritic silhouette of the ice neuron. Excitability adds an active cable. Field closure is the biophysical m\'elange in which the neuron helps write the field that acts back on it, sculpting its dynamics and, during growth, its structure. That is a biological neuron.\setcounter{mpfootnote}{1}\footnote{For Context \& overview see:\\ \url{https://neurovium.science/posts/pblog-IceNeuron}}
\end{abstract}

\maketitle

\pagestyle{fancy}
\fancyhf{}
\fancyhead[LE,RO]{\small Dendritic motifs: structure--function via field closure}
\fancyfoot[C]{\thepage}
\renewcommand{\headrulewidth}{0.4pt}

\section{Introduction}
\begin{figure*}[t]
\centering
\includegraphics[width=0.98\textwidth]{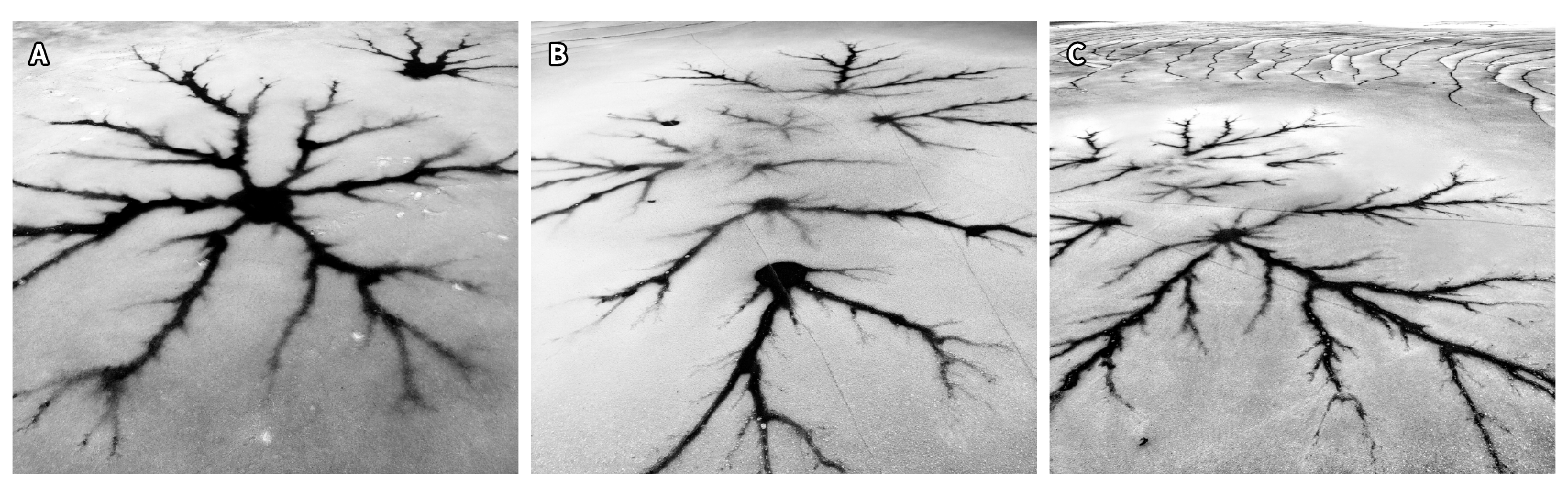}
\caption{\label{fig:ice}
Ice neurons. Melt channels in snow-covered ice on Walden Pond, Concord, Massachusetts, 1st of February 2020. Photographs by the author. (A)~Radial multipolar form around a central melt. (B)~A more polarized arbor with a soma-like pool. (C)~A field of competing trees beneath concentric ice waves. These silhouettes are the $\gamma=0$ object of the paper: a growth instability without an internal state that biological neurons posses.}
\end{figure*}

A frozen pond grows dark channels that look, at oak-leaf to a few-meters scale, like multipolar neurons, Purkinje cells, and pyramidal neurons [Fig.~\ref{fig:ice}; \cite{Pond}]. The resemblance is not merely a visual conceit. It is a statement about gradient-driven interfacial instability. A front that advances into a transported field---temperature in the ice, concentration in a solidification problem, voltage or morphogen in tissue---is morphologically unstable in a band of wavenumbers. The resulting morphological instability can be described at linear order by Mullins--Sekerka theory~\cite{MS64,Alexandrov24}. The nonlinear outcome is a tree, and related gradient-driven processes generate dendritic and fractal aggregates across the whole class of diffusion-limited growth models~\cite{Langer80,WittenSander81}.

Neuronal dendrites have been described in the same geometric language: fractal dimensions of reconstructed arbors~\cite{Caserta95,Smith21}, and a single wiring-cost rule that grows plausible trees across cell types~\cite{Cuntz10}. None of that is evidence that the tree was sculpted for the computing function that it enables. It is evidence that the tree is cheap.

The word \emph{motif} is asked to do two jobs in the scientific literature, and they are not the same. In one reading, a motif is a subgraph that occurs more often than in a randomized ensemble~\cite{Milo02,Alon07}. Growth rules produce such overrepresentation without selection on the subgraph~\cite{ValverdeSole05,Sole06}, in the architectural sense of Gould and Lewontin's spandrels~\cite{Gould79}: the space between two arches is a triangle because how the arches are put together not because sculpting a triangle was intended. In the other reading, a motif is a dynamical primitive---a feed-forward loop, an oscillator, a persistence filter---whose wiring is there because of what the wiring does~\cite{ShenOrr02,Alon07}. Ice neuron has the first kind of motif. It does not have the the second.

The distinction is easy to state and easy to lose once the resulting morphologies look alike. Temperature difference in the pond is not information difference in tissue, but both are gradients, and both sculpt a front. The important asymmetry is that a biological neuron is not only a passive consumer of its field: it can also write into that field, whereas an ice neuron has no corresponding independent internal state. Transmembrane ionic currents generate extracellular voltage gradients and electric field~\cite{Jefferys95,Buzsaki12,Petersen16,Bedard11}; that potential can feed back onto neighboring membrane by ephaptic coupling~\cite{Anastassiou11,GoldwynRinzel16,Pinotsis23}; and endogenous electric fields can guide cellular and neuritic growth~\cite{McCaig05}. The pattern is therefore capable, in tissue, of writing the field that continues to write the pattern. Ice does not do that with an independent internal state. Latent heat is slaved to the front velocity. There is no second variable on the interface that can oscillate when the front is held still.

In this paper, we write one continuum system that contains ice neuron as a limit and ask what finite source gain does. The morphological sector is the standard Mullins--Sekerka moving-front problem. The interfacial sector is a FitzHugh--Nagumo pair~\cite{FitzHugh61,Nagumo62}, chosen because it is the minimal internal state with a Hopf. The coupling is a single number $\gamma$ in the interfacial flux budget. At $\gamma=0$ one recovers a Mullins--Sekerka tree. At finite $\gamma$ the same tree, once frozen, retains activity longer after stimulation than the same cable at $\gamma=0$. That excess is the second reading of motif. The first reading never needed $\gamma$. The model below makes explicit how dendritic form and function interact to shape one another.

% The calculations below are numerical experiments: a cubic spectrum, a first-order resolvent formula, an integration on one grown graph, and an ensemble over eighty realizations of the growth rule. They are not a wet-lab measurement, and they are not a census over families of growth rules. Both absences are named in the Discussion.

\section{Model}

\subsection{Exterior field}

Let $\Omega(t)\subset\mathbb{R}^{2}$ be the condensed phase (ice, or the interior of a neurite in a crude identification) and $\Gamma=\partial\Omega$ its interface. In the exterior,
\begin{equation}
\partial_t\phi=D\nabla^2\phi.
\label{eq:diff}
\end{equation}

Quasi-static reduction, $\nabla^2\phi=0$, applies when diffusion relaxes the field rapidly compared with the motion of the interface, corresponding to the small-P'eclet-number limit. In this regime the elementary Mullins--Sekerka dispersion law takes the form $\omega\sim a|k|-b|k|^3$. We therefore keep diffusion in the statement of the model but use the quasi-static Dirichlet-to-Neumann map in the linear analysis, as is standard in the morphological-instability literature~\cite{MS64,Langer80}. This isolates the morphological instability from finite-P'eclet corrections.

Why use a single scalar field $\phi$? In the pond, $\phi$ is temperature. In tissue, the analogous quantity is extracellular potential, or more coarsely an ``activation field.'' The analogy between ice and tissue is not an identification of fields, but of dynamical roles: both contain a transported field $\phi$ that acts on an advancing interface; the distinction is whether an independent interfacial state can write back into that field. Finite $\gamma$ supplies precisely that feedback and allows the closure question to be posed within the same construction. One field is sufficient to pose that question. Two fields---temperature and voltage, or voltage and a morphogen---would refine the model, but are not prerequisites and would not change the underlying argument.

\subsection{Interfacial equilibrium and front motion}

The exterior field alone does not determine how the interface evolves. Two boundary relations close the morphological sector: one sets the value of the field at a curved interface, and the other converts the field gradient into motion of that interface. On $\Gamma$,

\begin{align}
\phi&=\Delta - d_0 \kappa,
\label{eq:GT}\\
v_n&=-K\,\mathbf{n}\!\cdot\!\nabla\phi\qquad(\gamma=0).
\label{eq:stef}
\end{align}

Equation~\eqref{eq:GT} is the Gibbs--Thomson relation. Curvature shifts the local interfacial equilibrium, so the capillary length $d_0$ penalizes strongly curved features and suppresses arbitrarily short-wavelength perturbations. It therefore supplies the ultraviolet cutoff of the Mullins--Sekerka instability. Equation~\eqref{eq:stef} is the Stefan flux balance in the absence of an additional interfacial source: it converts the normal gradient of the transported field into the normal velocity $v_n$ of the front. In solidification, this expresses conservation of heat across a moving phase boundary; in the geometric neurite analogy, it means that the front advances according to the local driving gradient. Section~\ref{sec:closure} will add the interfacial source proportional to $\gamma$; Eq.~\eqref{eq:stef} is displayed here first because it is the $\gamma=0$ law obeyed by the ice limit.

For a directional-growth phase diagram, one also needs a regime in which the planar front is linearly stable. A planar front with $d_0>0$ and no imposed stabilizing gradient is otherwise unstable at every undercooling. We therefore write the effective driving gradient as
\begin{equation}
G_{\mathrm{eff}} = G - G_c,
\label{eq:gradient}
\end{equation}
so that $G<G_c$ gives a linearly stable plane. Here $G_c$ is a phenomenological threshold representing the stabilizing gradient of a directional-solidification setting, rather than a microscopic parameter of the model. The motif argument does not depend on a particular value of $G_c$.

\subsection{Interfacial state}

Latent heat slaved to $v_n$ is not an internal degree of freedom:
if the front is frozen, that source vanishes. A neuron with a fixed
arbor, by contrast, remains dynamically active, and biological
dendrites are not passive geometric supports; individual branches
can behave as nonlinear computational subunits~\cite{Polsky04,LondonHausser05}.
The interface therefore requires an independent state whose dynamics
persist even when the front itself is held fixed.

The minimal membrane-like dynamical state capable of an intrinsic
Hopf instability is two-dimensional. FitzHugh--Nagumo
~\cite{FitzHugh61,Nagumo62} provides precisely that structure:
\begin{align}
\partial_t\psi&=\psi-\frac{\psi^3}{3}-w+\beta\,\phi\big|_{\Gamma}+D_\psi\partial_s^2\psi+I_{\mathrm{ext}},
\label{eq:fhn1}\\
\partial_t w&=\varepsilon\bigl(\psi-\gamma_w w+a\bigr).
\label{eq:fhn2}
\end{align}
Here $\psi$ is the fast interfacial variable and $w$ its slower
recovery variable. Tangential diffusion $D_\psi\partial_s^2\psi$
couples neighboring points along the interface and regularizes the
large-$k$ dynamics, while $I_{\rm ext}$ provides the probe used later
to measure persistence.

A Hodgkin--Huxley model~\cite{HH52} would introduce channel-level
biophysics and additional state variables without changing the
distinction being tested here. More importantly, a single leak
variable,
\[
\tau\dot{\psi}=-\psi+\beta\phi,
\]
cannot undergo a Hopf bifurcation. A delay can make such a scalar
feedback system oscillatory, but the oscillatory mechanism would then
reside in the imposed delay rather than in the interfacial state
itself. FitzHugh--Nagumo instead places the oscillatory instability
directly in the vector field on $\Gamma$, where the active membrane
dynamics reside.

\subsection{Field closure}
\label{sec:closure}

At $\gamma=0$, the interface is driven entirely by the gradient of the
exterior field: the state of the interface does not itself contribute to
that field. Field closure is introduced by allowing the interfacial
variable $\psi$ to contribute directly to the flux balance. We write this
contribution as a source density $\gamma\psi$, where the dimensionless
parameter $\gamma$ controls the strength of the feedback. The Stefan
condition therefore becomes
\begin{equation}
v_n=-K\bigl(\mathbf{n}\cdot\!\nabla\phi-\gamma\psi\bigr).
\label{eq:cl}
\end{equation}
Equation~\eqref{eq:stef} is recovered when $\gamma=0$. At finite $\gamma$, the motion
of the front is no longer determined solely by the externally established
gradient of $\phi$: the interfacial state also modifies the flux that
drives the front. Equivalently, rearranging the boundary condition, Neumann datum dictates
\begin{equation}
\mathbf{n}\cdot\nabla\phi
=
\gamma\psi-\frac{v_n}{K},
\label{eq:neumann}
\end{equation}
so that $\psi$ acts as a boundary source for the field. The coupling is therefore bidirectional. The field $\phi$ acts on the
interfacial state through the $\beta\phi|_\Gamma$ term in
Eq.~\eqref{eq:fhn1}, while the interfacial state $\psi$ writes back into the
field through the $\gamma\psi$ term in Eq.~\eqref{eq:cl}. Schematically,
\[
\phi
\;\xrightarrow{\;\beta\phi|_\Gamma\;}\;
\psi,
\qquad
\psi
\;\xrightarrow{\;\gamma\psi\;}\;
\phi.
\]
This reciprocal coupling is what we call field closure.

In the biological interpretation, the same structure corresponds to
ephaptic coupling~\cite{Jefferys95,Anastassiou11,GoldwynRinzel16,Pinotsis23}:
transmembrane currents generate extracellular potentials, and those
potentials in turn influence membrane dynamics. The parameter $\gamma$
does not represent a specific ion channel or microscopic coupling
mechanism; it measures the strength of this reciprocal field interaction
in the reduced model. Thus $\gamma=0$ gives the ice limit, in which the
front responds to a field without an independent interfacial state writing
back into it, whereas $\gamma>0$ introduces the additional feedback
available to an active neuronal interface.

We use an interfacial source rather than a bulk source because the active
degree of freedom represented by $\psi$ is localized on $\Gamma$. In the
biological reading, this corresponds to membrane-associated currents
rather than volumetric production throughout the surrounding medium. A
phase-field regularization would replace the sharp boundary source
$\gamma\psi$ by a smeared source distributed across a thin interfacial layer; the
sharp-interface formulation used here is the corresponding limiting
description.

\subsection{Isolated oscillator, parked subcritical}

Before examining the coupled front, we first locate the Hopf threshold
of the interfacial oscillator in isolation. Let $\phi_0$ be the base
value of $\phi$ on a planar front. A spatially uniform fixed point of
Eqs.~\eqref{eq:fhn1}--\eqref{eq:fhn2} satisfies
\[
w_0=\psi_0-\frac{\psi_0^3}{3}+\beta\phi_0,
\qquad
\psi_0-\gamma_w w_0+a=0.
\]
The local slope of the fast variable at that fixed point is
$f_\psi=1-\psi_0^2$. Linearizing the isolated $(\psi,w)$ dynamics
about this state gives the Jacobian
\begin{equation}
J_0=
\begin{pmatrix}
f_\psi & -1\\
\varepsilon & -\varepsilon\gamma_w
\end{pmatrix},
\label{eq:J0}
\end{equation}
with
\[
\operatorname{tr}J_0=f_\psi-\varepsilon\gamma_w,
\qquad
\det J_0=\varepsilon(\gamma_w f_\psi+1).
\]
For the two-dimensional system in Eq.~\eqref{eq:J0}, a Hopf
bifurcation occurs when $\operatorname{tr}J_0=0$ with
$\det J_0>0$, giving the isolated threshold
$f_\psi=\varepsilon\gamma_w$.

For the numerical calculations, we place the isolated oscillator \emph{below}
this threshold:
\[
f_\psi=0.06,\qquad
\varepsilon=0.10,\qquad
\gamma_w=1.15,
\]
so that $\operatorname{tr}J_0=-0.055$. This choice separates intrinsic
interfacial dynamics from effects introduced by field closure. If the
interfacial state were already oscillatory at $\gamma=0$, an
oscillatory response on the frozen tree could arise from the
FitzHugh--Nagumo dynamics alone and tells one nothing about field closure. By keeping the isolated oscillator
subcritical, any oscillatory instability that appears only at finite
$\gamma$ is instead a property of the coupled system and can be
attributed to the field-closure interaction, which distinguishes an ice neuron from a biological one.

\paragraph*{A note on what is not being claimed:}
The correspondence introduced above is deliberately coarse-grained: it
is a correspondence between dynamical roles, not a literal identification
of the physical constituents.

In our model, $\Omega$ is not cytoplasm, $\Gamma$ is not a lipid bilayer and the field
$\phi$ is not the membrane potential $V_m$. Likewise, the interfacial
source term is not intended as a microscopic model of any particular ion
channel or extracellular current pathway. The identification is instead
at the level of an internal dynamical state, a field that acts on the
interface, and a finite gain by which that state writes back into the
field.

\section{Linear theory}

\subsection{Planar front}

To characterize the onset of morphological instability, we linearize
about a uniformly translating planar front $x=Vt$. In the comoving frame,
a small transverse displacement of the interface is written in normal-mode
form as
\[
\zeta(y,t)=\hat{\zeta}\,e^{\omega t+iky}.
\]
In the quasi-static exterior, the corresponding perturbation of the field
decays away from the interface,
\[
\phi_1=Ae^{-|k|x}e^{\omega t+iky}.
\]
Applying the Gibbs--Thomson condition~\eqref{eq:GT} together with the
base-state gradient relates the field perturbation at the interface to
the displacement,
\begin{equation}
\phi_1\big|_{\Gamma}
=
(G_{\mathrm{eff}}-d_0 k^2)\hat\zeta.
\label{eq:phi-interface}
\end{equation}

Substituting Eq.~\eqref{eq:phi-interface} into the closure-modified
interfacial flux condition~\eqref{eq:cl} gives
\begin{equation}
\omega=p(k)\,|k|+K\gamma\,\chi(k,\omega),\qquad
p(k)=K(G_{\mathrm{eff}}-d_0 k^2),
\label{eq:disp-struct}
\end{equation}
where $\chi=\psi_1/\hat\zeta$ is the susceptibility of the interfacial
oscillator.

Equation~\eqref{eq:disp-struct} separates naturally into two contributions.
The term $p(k)|k|$ is the classical Mullins--Sekerka contribution to the
dispersion relation, while $K\gamma\chi(k,\omega)$ is the correction
introduced by field closure. At $\gamma=0$, the closure contribution
vanishes and the classical Mullins--Sekerka branch is recovered. Algebraic details are in the End Matter \ref{sec:appendix}.

Through Eq.~\eqref{eq:phi-interface}, the susceptibility $\chi$ itself
inherits the interfacial factor
$G_{\mathrm{eff}}-d_0k^2=p(k)/K$. Consequently, every closure contribution
in the linear theory is proportional to $p(k)\gamma$. This decomposition
is the central structural feature of the linear analysis: the
moving-front instability supplies the morphological branch, while field
closure couples that branch to the dynamics of the interfacial state.

\subsection{Interfacial susceptibility and coupled spectrum}

We next linearize the interfacial FitzHugh--Nagumo dynamics,
Eqs.~\eqref{eq:fhn1}--\eqref{eq:fhn2}, about the planar base state and
apply the same normal-mode perturbation used for the moving front. The
interfacial perturbations satisfy
\begin{align}
\omega\psi_1&=(f_\psi-D_\psi k^2)\psi_1-w_1+\beta\phi_1,
\label{eq:linpsi}\\
\omega w_1&=\varepsilon\psi_1-\varepsilon\gamma_w w_1,
\label{eq:linw}
\end{align}
Eliminating $w_1$ from Eqs.~\eqref{eq:linpsi}--\eqref{eq:linw} gives the
susceptibility of the interfacial state to the front displacement.
Substitution into Eq.~\eqref{eq:disp-struct}, followed by clearing the
denominator $\omega+\varepsilon\gamma_w$, yields the cubic characteristic
polynomial
\begin{equation}
\omega^3+A\omega^2+B\omega+C=0,
\label{eq:cubic}
\end{equation}
with coefficients given in Eq.~\eqref{eq:ABC} in the End Matter.

The uncoupled limit makes the spectral structure transparent. At
$\gamma=0$, Eq.~\eqref{eq:cubic} factors as
\begin{equation}
(\omega-p_k)\bigl[(\omega+\alpha)(\omega+\delta)+\varepsilon\bigr]=0.
\label{eq:factor}
\end{equation}
The first factor in Eq.~\eqref{eq:factor} reproduces the classical
Mullins--Sekerka growth mode, while the quadratic factor gives the
spectrum of the isolated FitzHugh--Nagumo state, including tangential
diffusion. Finite $\gamma$ couples these two sectors through the
coefficients $B$ and $C$ in Eq.~\eqref{eq:cubic}, so that the
morphological and interfacial modes no longer form independent spectral
branches. This spectral coupling is the linear signature of field
closure and provides the basis for the stability classification below.

\subsection{Classification}
\label{sec:classification}

For each $(G,\gamma)$, we evaluate all roots of Eq.~\eqref{eq:cubic}
over a grid of wavenumbers $k$ and identify the \emph{winning} root with the largest
real part. The state is then classified according to the character of
this dominant mode. It is \emph{planar} if
$\max \operatorname{Re}\omega\leq 0$, \emph{real-dominant} if the
winning root is real and positive, and \emph{complex-dominant} if the
winning root satisfies $\operatorname{Re}\omega>0$ and
$|\operatorname{Im}\omega|>0.05$. We use \emph{coexistent} for the case
in which the winning root is real while another complex root is also
unstable. The threshold on the imaginary part is a numerical convention
used to distinguish oscillatory roots from numerical proximity to the
real axis; it is not assigned a biological frequency scale.

The vocabulary matters. The presence of an unstable complex root does
not by itself mean that the fastest-growing morphological mode is
oscillatory. On the grid of Fig.~\ref{fig:spec}B the coexistent class
happens to be empty---wherever a complex root grows, it is also the
winner. Isolated FitzHugh--Nagumo never crosses at the parameters above.
Any complex-dominant cell is therefore a coupled-system cell, and the
word \emph{Hopf} is reserved for the planar-front tongue of
Fig.~\ref{fig:tongue}, where a growing complex root appears at parameters
for which the isolated oscillator is quiescent.

\section{A motif assay on a grown tree}

Linear theory on a plane says whether a front wants to wrinkle and
whether the wrinkle wants to ring. It does not say whether a finished
tree computes. To ask that question, we grow the tree as an ice neuron at
$\gamma=0$, freeze its geometry, and then ask what happens when finite
$\gamma$ is turned on, so that any additional dynamics arise from field
closure rather than from a change in morphology. The shape no longer
changes; only the dynamical closure does. That is the distinction that
turns the model of an ice neuron into that of a biological one.

\subsection{Growth rule}

Tips advance by a unit step with weak angular noise and a short-range repulsion off the existing skeleton, and split with probability $0.07$ into daughters at $\pm 0.62$~rad. This is not Laplacian growth and does not pretend to be. It is a growth rule in the sense of Sol\'e: local and memoryless except for self-avoidance~\cite{ValverdeSole05,Sole06}. The resulting polyline is resampled at $\Delta s=2.4$ into $N$ compartments ($N=117$ for the working seed) with a graph Laplacian $L$ (cable) and a screened ephaptic matrix
\begin{equation}
(G_{\mathrm{eph}})_{ij}=0.12\,\frac{e^{-r_{ij}/\lambda}}{r_{ij}}\qquad(i\neq j),\quad\lambda=6.
\label{eq:Geph}
\end{equation}
Screening is included because an unscreened $1/r$ interaction on a
hundred-node graph destabilizes the linear motif at vanishing $\gamma$
and erases the comparison we are trying to make. The End Matter \ref{sec:appendix} records the seeds and the discretization.

\subsection{Compartmental operator}

Once the grown arbor is frozen, we represent it as a compartmental graph.
Each compartment carries the local interfacial state $(\psi_i,w_i)$, while
the graph Laplacian $L$ couples neighboring compartments along the cable.
Collecting all compartmental variables into
$Y=(\boldsymbol{\psi},\boldsymbol{w})$, the linearized dynamics are
\begin{equation}
\dot{Y}=
\begin{pmatrix}
f_\psi I-g_cL+\beta\gamma G_{\mathrm{eph}} & -I\\
\varepsilon I & -\varepsilon\gamma_w I
\end{pmatrix}
Y+I_{\mathrm{ext}}(t),
\label{eq:tree}
\end{equation}
with $g_c=0.12$.

The two spatial operators represent different routes of interaction.
The Laplacian $L$ describes propagation constrained by the dendritic
cable, whereas $G_{\mathrm{eph}}$ couples compartments according to their
spatial separation rather than their adjacency on the graph. At
$\gamma=0$, the latter contribution vanishes, leaving a cable-coupled,
subcritical FitzHugh--Nagumo system---the compartmental counterpart of
classical cable theory~\cite{Rall62}. Finite $\gamma$ introduces an
additional nonlocal, distance-weighted interaction mediated by the
surrounding field. This term is the discrete counterpart of the
field-closure coupling introduced at the moving interface (Eq.~\eqref{eq:cl}) and provides
the graph-level representation of ephaptic coupling
~\cite{Anastassiou11,GoldwynRinzel16,Pinotsis23}.

\begin{figure*}[t]
\centering
\includegraphics[width=0.98\textwidth]{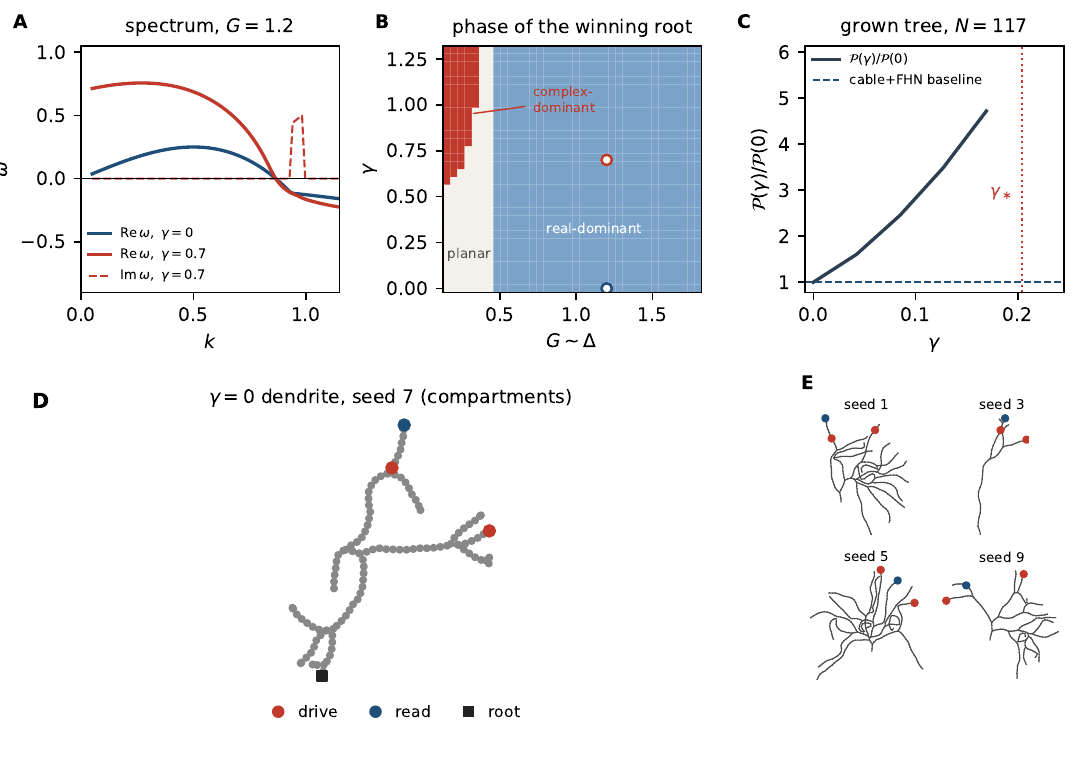}
\caption{\label{fig:spec}
(A)~Most unstable root of the cubic~\eqref{eq:cubic} at $G=1.2$. Isolated FitzHugh--Nagumo is subcritical. $\gamma=0.7$ lifts $\mathrm{Re}\,\omega$; a complex root appears only on the decaying side of the band.
(B)~Classification in $(G,\gamma)$ by the winning root: planar, real-dominant, complex-dominant (the coexistent class is empty on this grid). Ice is the line $\gamma=0$; circles mark the two spectra of (A).
(C)~Persistence fold-change on the tree in (D). The linear motif destabilizes at $\gamma_\ast\simeq 0.20$.
(D)~Self-avoiding tip-splitting dendrite grown at $\gamma=0$ (seed 7), $N=117$ compartments. Red: driven tips. Blue: readout. Square: root.
(E)~The same growth rule at four other seeds, with their own drive/read triples; the ensemble over eighty seeds is in Fig.~\ref{fig:ens}.}
\end{figure*}

\subsection{Persistence}

We next ask whether field closure changes how long activity remains
detectable on the frozen arbor after a transient input. Two distal tips
are driven by a rectangular pulse of duration $t_\ast=1.2$, and the
response is measured at a third distal tip. The three sites are chosen to
be well separated on the tree: among the twelve nodes farthest from the
root, the readout is selected to maximize the minimum pairwise distance
from the driven tips.

The persistence index
\begin{equation}
\mathcal{P}
=\frac{\int_{t_\ast}^{T}|\psi_{\mathrm{out}}|\,dt}{\int_0^{t_\ast}|I_{\mathrm{ext}}|\,dt}
\label{eq:P}
\end{equation}
measures the post-stimulus $L^1$ mass of the readout, normalized by the
total input. Integration is performed with forward Euler at
$\Delta t=0.02$ out to $T=40$. If the spectral abscissa of~(Eq.\eqref{eq:tree}) the
compartmental operator is nonnegative, the linear persistence  $\mathcal{P}$ is left
undefined, because the system has ceased to act as a stable filter.

The distal readout is deliberate. On a leaky cable of this length, a
distal pulse is strongly attenuated before reaching the soma at
$\gamma=0$, making the somatic response nearly zero ($\mathcal{P}(0)$ and any ratio $\mathcal{P}(\gamma)/\mathcal{P}(0)$ to its
baseline numerically ill-conditioned. A distal-to-distal assay avoids
that degeneracy while remaining a stringent test: transmission between
well-separated tips is weak when it must follow the cable alone. An
increase in persistence at finite $\gamma$ can therefore be attributed
to the additional field-mediated interaction rather than to a change in
the arbor itself.

\section{Results}

\subsection{Spectral deformation by field closure}

Figure~\ref{fig:spec}A shows how field closure reshapes the
linear spectrum at fixed $G=1.2$. The uncoupled case provides the
reference. At $\gamma=0$, Eq.~\eqref{eq:cubic} reduces through
Eq.~\eqref{eq:factor} to the usual real Mullins--Sekerka growth
band, with $\max\mathrm{Re}\,\omega\simeq0.25$ at
$k_\ast\simeq0.5$. Turning on field closure substantially lifts
this branch: at $\gamma=0.7$ the maximum reaches
$\mathrm{Re}\,\omega\simeq0.76$ and moves toward smaller $k$.
The change is therefore not only an increase in growth rate but
also a shift in the spatial scale selected by the instability.

The origin of that shift can already be read from the dispersion
relation. The bare morphological contribution in
Eq.~\eqref{eq:disp-struct} scales as $p(k)|k|$, whereas the
closure contribution enters through $p(k)\gamma$ without the
additional factor of $|k|$. Field closure consequently gives
greater relative weight to the long-wavelength part of the
spectrum, where the uncoupled Mullins--Sekerka rate itself tends
to zero. The effect is not a uniform spectral amplification of the Mullins--Sekerka
band: field closure warps the dispersion curve, lifting its
long-wavelength side more strongly and shifting the selected instability
mode toward smaller $k$.
% The coupled spectrum is therefore pulled upward and
% toward smaller $k$, rather than being a simple vertical
% rescaling of the $\gamma=0$ band.

Oscillatory character appears only later in $k$. Near
$k\gtrsim1$, on the decaying side of the real branch, the
selected root acquires a nonzero imaginary part. It does not,
however, overtake the long-wavelength real maximum. Thus this
particular $(G,\gamma)$ remains real-dominant in the sense of
Sec.~\ref{sec:classification}: field closure has strongly
reorganized the morphological spectrum, while the oscillatory
sector is present but is not yet the instability selected by the
front.
\begin{figure*}[t]
\centering
\includegraphics[width=0.98\textwidth]{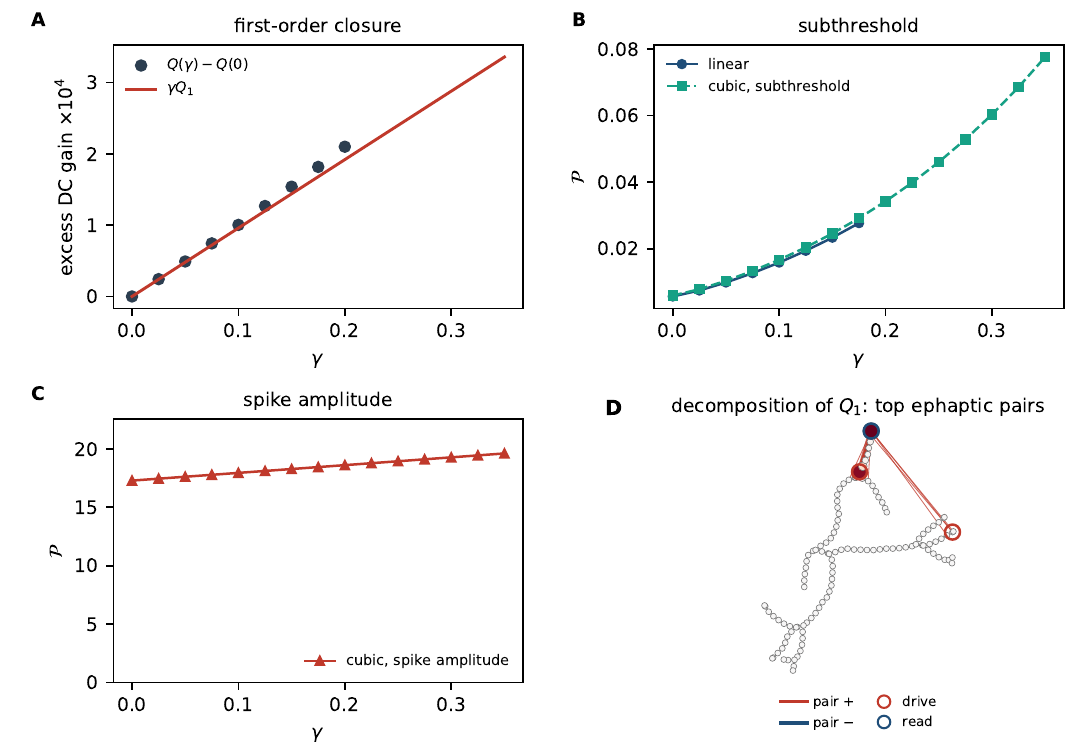}
\caption{\label{fig:enrich}
(A)~Excess DC gain on the frozen tree versus the first-order law $\gamma Q_1$.
(B)~Subthreshold cubic FitzHugh--Nagumo tracks linear $\mathcal{P}$ and does not terminate at $\gamma_\ast$.
(C)~Spike-amplitude cubic FitzHugh--Nagumo: the cable already transmits; $\gamma$ is a fractional correction.
(D)~Spatial decomposition of $Q_1=\sum_{ij}z_{\mathrm{out},i}B_{ij}z_{\mathrm{in},j}$: line thickness is the magnitude of a pair's contribution (red positive, blue negative; forty largest pairs shown), node color and size the nodewise sum. Open circles: drive (red) and read (blue). The motif's work is done by a few direct field shortcuts, not by the silhouette.}
\end{figure*}

\subsection{Phase structure of the coupled front}

Figure~\ref{fig:spec}B extends the spectral classification of
Sec.~\ref{sec:classification} across the $(G,\gamma)$ plane.
Along the ice axis, $\gamma=0$, the familiar transition is recovered:
for $G\lesssim G_c$ the planar front is stable, whereas above $G_c$
the Mullins--Sekerka instability is selected and the system is
real-dominant. Finite field closure changes the low-$G$ sector.
As $\gamma$ increases, part of the otherwise planar region is replaced
by a complex-dominant instability, while the higher-$G$ sector remains
controlled by the real morphological mode. The complex-dominant region
therefore does not represent oscillations superposed on an already
selected growing mode; it opens where the uncoupled morphological band
would still be closed.

On this grid, the complex-dominant region contains 72 of 1152 cells,
and no cell is classified as coexistent. Because the isolated
FitzHugh--Nagumo pair remains below its Hopf threshold throughout,
every complex-dominant cell is a property of the coupled system:
finite $\gamma$ has made a complex root both unstable and spectrally
dominant. The distinction is especially clear along $\gamma=0$.
Ice neurons can occupy either the planar or real-dominant sectors as
$G$ is varied, but they have no route into the complex-dominant sector.
In that precise linear sense, the ice neurons in Fig.~\ref{fig:ice}
are silhouettes. In that sense, Fig.~\ref{fig:spec}B separates morphological instability from
the additional dynamical instability made available by field closure.

\subsection{Persistence on the grown tree}

We next ask whether the same closure parameter that reshapes the planar
spectrum changes what a completed dendrite does once its geometry is
held fixed. Figure~\ref{fig:spec}C shows the normalized persistence
$\mathcal{P}(\gamma)/\mathcal{P}(0)$, with $\mathcal{P}$ defined in
Eq.~\eqref{eq:P}, on the working tree. The $\gamma=0$ system provides
the active-cable baseline: it is linearly stable, with spectral
abscissa $-0.028$, and distal-to-distal transmission is small but
finite, $\mathcal{P}\simeq5.6\times10^{-3}$. As field closure is
increased, persistence rises monotonically, reaching roughly $4.7$
times its baseline value by $\gamma\simeq0.17$. The enhancement
therefore develops while the linear system remains stable; it is not
simply the consequence of crossing an instability.

That stable window has a finite endpoint. The spectral abscissa of the
operator in Eq.~\eqref{eq:tree} approaches zero with increasing
$\gamma$ and crosses at $\gamma_\ast\simeq0.20$. Beyond
$\gamma_\ast$ the linear system no longer defines a stable persistence
filter, so linear $\mathcal{P}$ is not reported. The relevant feature
of Fig.~\ref{fig:spec}C is thus the rise preceding the pole: field
closure progressively increases the response retained at a distant
readout before destabilizing the linear dynamics.

The geometry underlying every point in this curve is unchanged.
Figure~\ref{fig:spec}D shows that frozen $\gamma=0$ tree, with
$N=117$ compartments and the two driven tips and distal readout used
for the assay. Thus the increase in Fig.~\ref{fig:spec}C cannot be
attributed to a change in branching or path length: the graph is fixed,
and only the field-coupling gain has changed. The rise below
$\gamma_\ast$ is therefore the dynamical motif being assayed here,
rather than a new statistic of the dendritic silhouette.
Figure~\ref{fig:spec}E places the working tree alongside four
additional realizations of the same growth rule, each with its own
extremal drive/read triple, anticipating the ensemble test below.

\subsection{First-order excess}

The monotonic rise in persistence suggests that field closure is adding
a systematic contribution to transmission rather than merely moving the
system toward its instability. To isolate that contribution, we write
$A(\gamma)=A_0+\gamma B$, with $B$ the ephaptic block
$\beta G_{\mathrm{eph}}$ acting on $\psi$. For a stable Hurwitz
operator, the integrated linear response to a unit-area impulse is the
DC gain
\begin{equation}
Q(\gamma)
=-\mathbf{e}_{\mathrm{out}}^{\top}
A(\gamma)^{-1}\mathbf{e}_{\mathrm{in}}.
\label{eq:Q}
\end{equation}
Expanding the resolvent about the active-cable limit,
$A^{-1}=A_0^{-1}-\gamma A_0^{-1}BA_0^{-1}+O(\gamma^2)$, gives
\begin{equation}
Q(\gamma)-Q(0)=\gamma Q_1+O(\gamma^2),\qquad
Q_1=
\mathbf{e}_{\mathrm{out}}^{\top}
A_0^{-1}BA_0^{-1}\mathbf{e}_{\mathrm{in}}.
\label{eq:Q1}
\end{equation}
Thus the leading effect of field closure is summarized by a single
kernel $Q_1$ attached to the chosen graph, drive, and readout. If
$Q_1$ vanishes, closure does no first-order motif work on that triple;
its sign and magnitude otherwise specify the direction and strength of
the leading correction.

For the working tree, $Q_0=3.06\times10^{-7}$ and
$Q_1=9.60\times10^{-4}$. Figure~\ref{fig:enrich}A shows that the
measured excess $Q(\gamma)-Q(0)$ follows the first-order prediction
$\gamma Q_1$ closely, with mean relative error $0.027$ over
$0<\gamma\le0.12$. Although the persistence $\mathcal{P}$ is not
identical to $Q$---it measures the post-pulse $L^1$ mass of
$|\psi|$---both probe the same linear operator. The increase seen in
the persistence assay therefore has a perturbative counterpart whose
leading coefficient is calculated directly from the operator rather
than fitted to the response.

The same kernel also identifies where that first-order effect is
carried. Writing
$Q_1=z_{\mathrm{out}}^{\top}Bz_{\mathrm{in}}
=\sum_{ij}z_{\mathrm{out},i}B_{ij}z_{\mathrm{in},j}$, with
$A_0z_{\mathrm{in}}=\mathbf{e}_{\mathrm{in}}$ and
$A_0^{\top}z_{\mathrm{out}}=\mathbf{e}_{\mathrm{out}}$, assigns a
signed contribution to every ephaptic pair $(i,j)$.
Figure~\ref{fig:enrich}D maps these contributions back onto the frozen
tree. They are highly localized rather than spread over the
silhouette: on this realization, the forty largest pairs account for
$99.9\%$ of the summed absolute contribution and form direct,
non-cable shortcuts linking the driven regions to the readout. The
first-order excess is therefore not only computable as a scalar; its
field-mediated support can be identified on the tree itself.

\subsection{Nonlinear persistence: subthreshold and regenerative regimes}

The linear persistence assay terminates at $\gamma_\ast$ because the
spectral abscissa of the linear operator crosses zero. That pole need
not imply a divergence of the underlying nonlinear dynamics. To test
what survives beyond the linear approximation, we restore the cubic
FitzHugh--Nagumo vector field in deviations
$u=\psi-\psi_0$,
\begin{equation}
\partial_t u
=
f_\psi u-\psi_0u^2-\frac{u^3}{3}-v
+\bigl(-g_cL+\beta\gamma G_{\mathrm{eph}}\bigr)u+I,
\label{eq:cubic-u}
\end{equation}
with $\psi_0=-\sqrt{1-f_\psi}$.

The answer depends strongly on drive amplitude.
For a subthreshold pulse of amplitude $0.05$,
Fig.~\ref{fig:enrich}B shows that the cubic response follows the
linear persistence closely throughout their common stable window.
At $\gamma_\ast$ the two descriptions part company: the linear
response ceases to be defined, whereas the nonlinear response remains
finite and continues smoothly as $\gamma$ is increased. The linear
pole is therefore a property of the linearization, not a divergence of
the nonlinear system. In this regime, field closure remains a major
contributor to distal transmission.

The situation is different for a spike-sized pulse of amplitude $0.8$.
As Fig.~\ref{fig:enrich}C shows, the $\gamma=0$ active cable already
produces a large response, $\mathcal{P}\simeq17$, indicating that a
regenerative event can propagate across the tree without field
closure. Increasing $\gamma$ still enhances that response, but only
from about $17$ to $19.6$ by $\gamma=0.35$. Field closure therefore
plays different roles in the two amplitude regimes: it strongly
enhances weak distal transmission below threshold, whereas once the
cable itself supports a regenerative event, it becomes a comparatively
smaller correction to an already large response.

The motif therefore survives in both amplitude regimes, but with
different weights. Below threshold, with transmission nearly absent at
$\gamma=0$, closure is almost the transmission. At spike amplitude,
closure is a correction to a cable spike that ice still does not have,
because ice has no cubic.

\subsection{A Hopf tongue below the isolated threshold}

The complex-dominant region of Fig.~\ref{fig:spec}B shows that field
closure can produce an oscillatory instability, but it does not by
itself separate that effect cleanly from morphological growth. At
$G=1.2>G_c$, the Mullins--Sekerka band is already open and a cut
through $(\gamma,\operatorname{tr}J_0)$ is dominated by the real
growth mode. To isolate the oscillatory effect, we instead move below
the morphological threshold and set $G=0.25<G_c$, so that the
uncoupled planar front is stable.

\begin{figure}[t]
\centering
\includegraphics[width=\columnwidth]{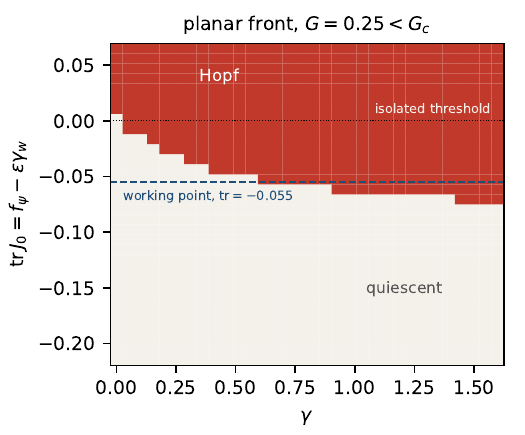}
\caption{\label{fig:tongue}
Hopf tongue on a planar front ($G=0.25<G_c$). Dotted: isolated threshold $\mathrm{tr}\,J_0=0$. Dashed: the working value $\mathrm{tr}=-0.055$. Red: a growing complex root exists. Cream: quiescent.}
\end{figure}

Figure~\ref{fig:tongue} then varies the field-closure gain against the
trace of the isolated FitzHugh--Nagumo Jacobian in
Eq.~\eqref{eq:J0}. At $\gamma=0$, the familiar isolated threshold is
recovered: oscillatory growth occurs only when
$\operatorname{tr}J_0\geq0$. Finite $\gamma$ bends that boundary
downward. As closure is strengthened, a growing complex root of the
coupled cubic, Eq.~\eqref{eq:cubic}, appears for increasingly negative
values of $\operatorname{tr}J_0$, opening an oscillatory region inside
parameter space where the isolated oscillator remains quiescent. On
the scanned grid, this extension below the isolated threshold contains
176 cells.

The working oscillator makes the effect concrete. Its
$\operatorname{tr}J_0=-0.055$ lies below the isolated Hopf threshold,
so at $\gamma=0$ it sits in the quiescent region. Following the dashed
line horizontally across Fig.~\ref{fig:tongue}, sufficiently strong
field closure carries that same oscillator across the coupled boundary
and into the Hopf region without changing its isolated dynamics.
The oscillator has not been moved through its own Hopf threshold;
the threshold of the coupled system has moved around it. That is the
$\gamma$-driven Hopf.

\subsection{Robustness across the growth ensemble}
\label{sec:ensemble}

\begin{figure*}[t]
\centering
\includegraphics[width=0.98\textwidth]{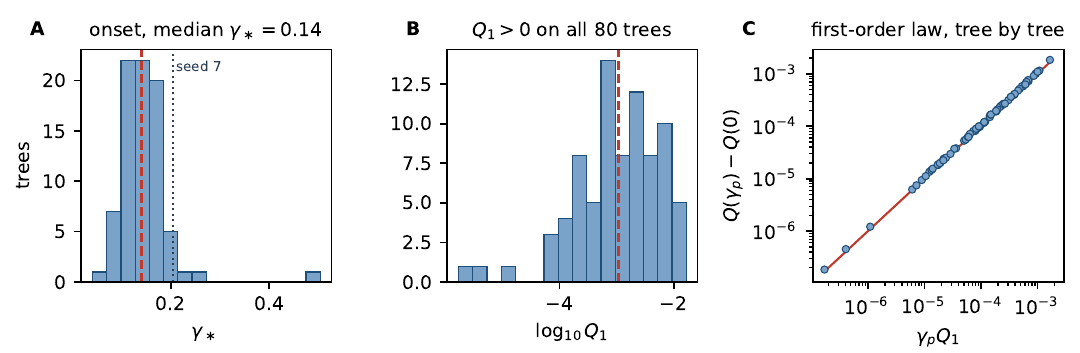}
\caption{\label{fig:ens}
Robustness over eighty growth seeds (End Matter, Sec.~\ref{sec:em-ensemble}).
(A)~Onset $\gamma_\ast$ of the frozen-tree instability; dashed: ensemble median, dotted: the working tree.
(B)~First-order kernel $Q_1$; positive on every tree.
(C)~First-order law tree by tree: excess DC gain at the per-tree probe $\gamma_p=\min(0.1,\,0.8\gamma_\ast)$ against $\gamma_p Q_1$; the diagonal is not a fit.}
\end{figure*}

The tree of Fig.~\ref{fig:spec}D is an ensemble of one.
Figure~\ref{fig:ens} repeats the construction over $80$ independent
seeds of the same growth rule, with $N=29$--$358$ compartments
(median $226$) and the extremal drive/read triple re-selected on each
tree. Three statements survive the ensemble.

First, the kernel has a sign: $Q_1>0$ on all eighty trees, spanning
$1.7\times10^{-6}$ to $1.7\times10^{-2}$ with median
$1.1\times10^{-3}$. The magnitude is geometry-dependent, but the
direction of the first-order closure effect is not.

Second, the stable linear window is generic but bounded.
The onset $\gamma_\ast$ has median $0.14$, with interquartile range
$0.12$--$0.17$ and extremes $0.04$ and $0.50$.
The working tree, at $\gamma_\ast\simeq0.20$, sits toward the upper
part of the distribution but is not exceptional.

Third, the first-order law of Eq.~\eqref{eq:Q1} is not a property of
the working seed. Probing each tree within its own stable window at
$\gamma_p=\min(0.1,\,0.8\gamma_\ast)$, the exact excess
$Q(\gamma_p)-Q(0)$ follows $\gamma_pQ_1$ across four decades, with
median relative deviation $0.06$ [Fig.~\ref{fig:ens}C]. The diagonal
is not a fit; each prediction comes from that tree's own
$\gamma=0$ operator.

The persistence fold itself is deliberately not promoted to an
ensemble statistic. Among the trees whose stable window reaches
$\gamma=0.1$ (69 of 80 trees) , $\mathcal{P}(0.1)/\mathcal{P}(0)>1$ on every tree,
but its magnitude spans many orders because the $\gamma=0$
distal-to-distal baseline collapses with tree depth
($Q(0)$ ranges from $3\times10^{-54}$ to $2\times10^{-4}$).
On a deep tree the fold mostly measures the death of its own
denominator. The ensemble-stable statements are therefore the sign of
$Q_1$ and the first-order law, and they both \emph{hold} on every tree generated by
this growth rule.

\section{Discussion}

The model set out to keep a resemblance and lose an identification. Ice neurons and biological neurons can look like the same object because they are products of the same kind of gradient instability [Fig.~\ref{fig:ice}]. They are not the same object because biological neurons can write the
extracellular field and be acted on by that field through ephaptic
coupling~\cite{Anastassiou11,Pinotsis23}. What the present calculation adds is that,
once this reciprocity is made explicit through finite $\gamma$, it
closes a dynamical loop that ice does not possess.

That distinction is the link to the two ``seemingly contrasting'' readings of motifs \cite{Alon07,Sole06}. Sol\'e's critique of motif census is correct for any ensemble generated by a growth rule~\cite{ValverdeSole05,Sole06}. The $\gamma=0$ trees of Fig.~\ref{fig:spec} are such an ensemble. Counting three-tip subgraphs on them, or on pond ice~\cite{Pond}, or on a Cuntz arbor~\cite{Cuntz10}, is a statement about how the object was grown. Gould and Lewontin~\cite{Gould79} already had the language in their portrayal of \emph{``spandrels''}: the triangle is there because the arches are there. The functional reading associated with Alon's motif program~\cite{Milo02,ShenOrr02,Alon07} is a different claim: that a recurrently observed wiring pattern constitutes a dynamical primitive because of what it does when driven. On the frozen tree, that claim is $Q(\gamma)-Q(0)=\gamma Q_1+O(\gamma^2)$ below threshold---with the kernel now localized to specific field shortcuts [Fig.~\ref{fig:enrich}D] and positive across the growth ensemble [Fig.~\ref{fig:ens}]---and a cubic correction at spike amplitude. It is empty at $\gamma=0$ in the linearization. It is not empty at $\gamma=0$ once the interface can spike---but that spike is the internal state that the ice neuron does not have.

\begin{table}[b]
\caption{\label{tab:hierarchy}Three possessions, not one. Each row adds a property the previous row lacks; the census of silhouettes cannot distinguish the rows.}
\begin{ruledtabular}
\begin{tabular}{lccc}
 & growth & internal & field \\
 & structure & dynamics & feedback \\
\hline
ice & \checkmark & -- & -- \\
active cable ($\gamma=0$) & \checkmark & \checkmark & -- \\
field-closed neuron ($\gamma>0$) & \checkmark & \checkmark & \checkmark \\
\end{tabular}
\end{ruledtabular}
\end{table}

The same distinction has a physical-computation reading. Neuronal computation is implemented in physical degrees of freedom: membrane state, ionic currents, cable propagation, and extracellular fields are parts of the substrate, not observer-side labels placed on it. A categorical account of physical computation makes this explicit by requiring correspondence between physical processes and computational operations, rather than identifying arbitrary physical evolution with computation~\cite{DehghaniCaterina24}. A stronger closure criterion then separates ordinary physical feedback from autonomous physical computation: internally realized readout states must participate in selecting subsequent operations~\cite{Dehghani26}. Field closure in the present model is deliberately the weaker, dynamical statement. The membrane state writes $\phi$, and $\phi$ acts back on subsequent membrane dynamics; ephaptic coupling supplies a biologically established mechanism for exactly such a loop~\cite{Anastassiou11,GoldwynRinzel16}. Whether that loop additionally realizes computational closure in the stronger sense depends on the readout and transition structure carried by the resulting dynamics, not on feedback alone. Table~\ref{tab:hierarchy} therefore provides a hierarchy: growth supplies structure; an internal state supplies autonomous dynamics; $\gamma$ supplies the physical loop; and computation, in the strong sense, is a further question about what the loop implements.

The calculations are numerical experiments on one growth rule (eighty seeds of it). They are not extracellular recordings, and they are not a comparison of diffusion-limited, wiring-cost, and pond-ice skeletons. Those are the next measurements, not missing code.

Several limitations are load-bearing and should be read as such.

First, the selected morphological mode in Fig.~\ref{fig:spec}A remains real at the plotted $(G,\gamma)$. The complex-dominant region in Fig.~\ref{fig:spec}B and the tongue in Fig.~\ref{fig:tongue} are real, but they are not the same statement as ``the growing tip rings.'' A growing-tip Hopf is a prediction for a different cut through parameter space, or for a nonlinear amplitude equation that this paper does not write.

Second, the growth rule is not Laplacian growth. An honest Mullins--Sekerka dendrite at $\gamma=0$ would be a better geometry on which to sit the motif assay. The self-avoiding splitter was used because it is cheap, reproducible, and already in the Sol\'e class. Replacing it with a phase-field or a pinched Laplacian-growth tree does not change the logic; it changes $L$ and $G_{\mathrm{eph}}$.

Third, linear $\mathcal{P}$ hits a pole at $\gamma_\ast$. The cubic on the same graph does not. Subthreshold pulses follow~\eqref{eq:Q1}; spike-sized pulses ride the cable and treat $\gamma$ as a perturbation. Both are reported because they are different sentences about the same $\gamma$.

Fourth, $\phi$ in tissue is not a single scalar obeying~\eqref{eq:diff} in an unbounded exterior. Extracellular space is a thin, tortuous conductor~\cite{NicholsonSykova98}. The Green function that replaces $e^{-r/\lambda}/r$ is a problem in its own right~\cite{Jefferys95,Anastassiou11,Pinotsis23}, and cable-theoretic frameworks in which the generated $V_e$ feeds back on $V_m$ show what a more realistic closure operator looks like~\cite{GoldwynRinzel16}. The present $G_{\mathrm{eph}}$ is a placeholder with one screening length. What it has to get right is topology: a coupling that does not follow the cable graph.

Fifth, no synapse is in the model. That is deliberate. The question was whether field closure, by itself, turns a growth-rule tree into a dynamical motif. Adding synapses would add Alon's original substrate and make the comparison to ice less sharp.

\paragraph*{What would withdraw the claim:} If $\mathcal{P}(\gamma)$ on a frozen $\gamma=0$ tree were independent of $\gamma$ below the first instability---on the working tree, or on the ensemble of Sec.~\ref{sec:ensemble}---field closure would not be doing motif work in this reduction, and the model would be a linear instability calculation about dendrites that look like ice. The cubic and the phase diagram would remain. The conclusion about biological dendritic structure seen as a dynamical motif would not.

\paragraph*{What the claim does not withdraw:} Subgraph overrepresentation on grown trees can still be a spandrel after every number in Sec.~V. The two readouts were never in competition for the same observable. They become competitors only when a silhouette is misread—either as proof that (dendritic) form must be necessary for function, or as grounds to deny function because the same form can arise without it. Pond ice is the existence proof that this inference fails [Fig.~\ref{fig:ice}; \cite{Pond}]. Finite $\gamma$ is the existence proof that it need not fail for a biological neuron.

The modeling scope is intentionally focused. If the object of study is a pattern, Mullins--Sekerka and a growth rule would suffice, and Sol\'e is the right reading of any motif one then counts. If the object of study is a function sitting on that pattern, the pattern has to source the field that drives it, and the interface has to have a state that is not slaved to the normal velocity. That is the functional content of $\gamma$ and of $(\psi,w)$. The same test extends to NeuroAI and design of bio-inspired networks: architectures that borrow dendritic geometry---dendrite-inspired networks and neuromorphic dendritic substrates~\cite{Baek24}---inherit only the first row of Table~\ref{tab:hierarchy}, and acquire the third only when their units write into a shared physical field that acts back on the computation, the closure that physical reservoirs can implement~\cite{Dehghani26} and that a copied silhouette cannot. The calculations show that those two ingredients are sufficient to move $\mathcal{P}$, to localize where they move it, and to open an oscillatory region that ice can not enter. \emph{Morphology requires a growth rule; dynamics requires an internal state; autonomous computation requires closure}.

% ----------------------------------------------------------------------
% Acknowledgement
% ----------------------------------------------------------------------
\begin{acknowledgments}
This study traces its origins to a winter walk, when I encountered striking ``ice neuron'' motifs along the surface of a frozen pond. Only later did I learn of neurologist Sandra Kostyk's earlier report describing similar formations, which she termed ``pond neurons''~\cite{Pond}. Although such structures would typically be regarded as incidental spandrels of ice and branching geometry, the patterns prompted a different question: what additional physical principle distinguishes a neuronal dendrite from a geometrically similar growth structure? The framework developed here grew from the possibility that field closure and ephaptic organization provide part of that distinction, and from the effort to understand how electric fields and dendritic geometry co-shape one another. I acknowledge the unusual circumstances that afforded the observation and the clarity for these reflections.
\end{acknowledgments}

% ----------------------------------------------------------------------
% Data/Code
% ----------------------------------------------------------------------

\section*{Code Availability}
Code implementing the model, the growth rule, the linear and cubic integrations, and the eighty-seed ensemble is available at: \\ \url{https://github.com/neurovium/IceNeuron}.

% ----------------------------------------------------------------------
% Appendix
% ----------------------------------------------------------------------

\appendix
\section{}
\label{sec:appendix}
\subsection{Fixed point and isolated Hopf}

Let $\phi_0$ be the base interfacial potential. A uniform fixed point of~\eqref{eq:fhn1}--\eqref{eq:fhn2} satisfies $w_0=\psi_0-\psi_0^3/3+\beta\phi_0$ and $\psi_0-\gamma_w w_0+a=0$. The $\psi$-linearization coefficient is $f_\psi=1-\psi_0^2$. Isolated Hopf at $\mathrm{tr}\,J_0=0$ with $\det>0$, i.e.\ $f_\psi=\varepsilon\gamma_w$. Working values: $f_\psi=0.06$, $\varepsilon=0.10$, $\gamma_w=1.15$, so $\mathrm{tr}=-0.055$ (subcritical). $\beta=1$, $D_\psi=0.20$. The rest-state used in the cubic integration is $\psi_0=-\sqrt{1-f_\psi}\simeq -0.9695$.

\subsection{Closure and cubic coefficients}

Write $\phi_1=Ae^{-|k|x}e^{\omega t+iky}$ in the exterior frame of a front $x=Vt+\zeta$. Gibbs--Thomson plus the base gradient give $\phi_1|_{\Gamma}=(G_{\mathrm{eff}}-d_0k^2)\hat\zeta=(p/K)\hat\zeta$. The normal gradient of the perturbation at the front is $-|k|A$, so the modified Stefan law~\eqref{eq:cl} linearizes to
\begin{equation}
\omega\hat\zeta=K\bigl(|k|A+\gamma\psi_1\bigr),
\qquad\text{i.e.}\qquad
\omega=p(k)|k|+K\gamma\,\chi,
\label{eq:em-disp}
\end{equation}
with $\chi=\psi_1/\hat\zeta$ and $p(k)=K(G_{\mathrm{eff}}-d_0k^2)$. From linearized FitzHugh--Nagumo with tangential diffusion,
\begin{align}
\omega\psi_1&=(f_\psi-D_\psi k^2)\psi_1-w_1+\beta\phi_1,\\
\omega w_1&=\varepsilon\psi_1-\varepsilon\gamma_w w_1,
\end{align}
eliminating $w_1$ and $\phi_1$ gives $\chi=\beta(p/K)\bigl[\omega+\alpha+\varepsilon/(\omega+\delta)\bigr]^{-1}$: the susceptibility carries the Gibbs--Thomson factor $p/K$, which is why every closure term below is proportional to $p(k)\gamma\beta$. Clearing denominators produces~\eqref{eq:cubic} with
\begin{align}
\alpha&=-f_\psi+D_\psi k^2,\qquad
\delta=\varepsilon\gamma_w,\qquad
s=\alpha+\delta,
\nonumber\\
q&=\alpha\delta+\varepsilon,\qquad
p_k=p(k)|k|,
\nonumber\\
A&=s-p_k,
\nonumber\\
B&=q-p_k s-p(k)\gamma\beta,
\nonumber\\
C&=-p_k q-p(k)\gamma\beta\,\delta.
\label{eq:ABC}
\end{align}
At $\gamma=0$ the cubic factors as $(\omega-p_k)\bigl[(\omega+\alpha)(\omega+\delta)+\varepsilon\bigr]=0$: Mullins--Sekerka mode times isolated FitzHugh--Nagumo. Roots are the eigenvalues of the companion matrix. Classification in Fig.~\ref{fig:spec}B follows Sec.~\ref{sec:classification} on a $k$-grid.

\subsection{First-order kernel}

On the frozen graph, $A(\gamma)=A_0+\gamma B$ with $B$ the block $\beta G_{\mathrm{eph}}$ acting on $\psi$ and zeros on $w$. If $A_0$ is Hurwitz,
\begin{equation}
Q(\gamma)=-\mathbf{e}_{\mathrm{out}}^{\top}A(\gamma)^{-1}\mathbf{e}_{\mathrm{in}}
=Q(0)+\gamma Q_1+O(\gamma^2),
\end{equation}
with $Q_1=\mathbf{e}_{\mathrm{out}}^{\top}A_0^{-1}B A_0^{-1}\mathbf{e}_{\mathrm{in}}$. Equivalently $Q_1=z_{\mathrm{out}}^{\top}B z_{\mathrm{in}}$ where $A_0 z_{\mathrm{in}}=\mathbf{e}_{\mathrm{in}}$ and $A_0^{\top}z_{\mathrm{out}}=\mathbf{e}_{\mathrm{out}}$. The map in Fig.~\ref{fig:enrich}D plots the symmetrized pair terms $(z_{\mathrm{out},i}B_{ij}z_{\mathrm{in},j}+z_{\mathrm{out},j}B_{ji}z_{\mathrm{in},i})$ for the forty pairs of largest magnitude, and colors each node by its summed contribution. On the working tree, $Q_0=3.058\times 10^{-7}$ and $Q_1=9.600\times 10^{-4}$.

\subsection{Grown tree}

Tips advance by a unit step with angular noise of width $0.18$ and a short-range $1/r^3$ repulsion inside radius $4$, and split with probability $0.07$ into $\pm 0.62\,\mathrm{rad}$ daughters (NumPy generator, working seed $7$). Polylines are resampled at $\Delta s=2.4$ into $N$ compartments ($N=117$ at seed 7). Cable coupling is the graph Laplacian with $g_c=0.12$. Ephaptic coupling is~\eqref{eq:Geph}. The $2N\times 2N$ linear operator is~\eqref{eq:tree}. Drive two distal tips ($t\le 1.2$), read a third chosen to maximize pairwise separation among the twelve farthest nodes from the root. Linear $\mathcal{P}$ uses $\Delta t=0.02$ forward Euler out to $t=40$. Cubic integration of~\eqref{eq:cubic-u} uses $\Delta t=0.01$. The first linear pole crosses at $\gamma_\ast\simeq 0.204$ (bisection on the spectral abscissa).

\subsection{Ensemble}
\label{sec:em-ensemble}

Seeds $1$--$80$ of the same growth rule and discretization, drive/read triples re-selected per tree by the same extremal rule. Per tree: $Q_0$, $Q_1$ from the resolvent formulas above; $\gamma_\ast$ by bisection on the spectral abscissa of~\eqref{eq:tree}; the first-order law tested at $\gamma_p=\min(0.1,\,0.8\gamma_\ast)$. Tree sizes span $N=29$--$358$ (median $226$). Results: $Q_1>0$ on $80/80$ trees (median $1.1\times10^{-3}$, range $1.7\times10^{-6}$--$1.7\times10^{-2}$); $\gamma_\ast$ median $0.14$, interquartile range $0.12$--$0.17$, extremes $0.04$ and $0.50$; median relative error of $\gamma_pQ_1$ against the exact excess $0.059$ (maximum $0.105$). Where $\gamma_\ast>0.1$ (69 trees), $\mathcal{P}(0.1)/\mathcal{P}(0)>1$ on every tree; the fold magnitude is dominated by the depth-dependence of the $\gamma=0$ baseline ($Q_0$ range $3\times10^{-54}$--$2\times10^{-4}$) and is reported only as a sign.

\subsection{Parameters}

Morphology: $K=d_0=\beta=1$, $G_c=0.45$, $G=1.2$ in Fig.~\ref{fig:spec}A and $G=0.25$ in Fig.~\ref{fig:tongue}. FitzHugh--Nagumo as in A.1. Tree as in A.4. Code: \texttt{tree\_model.py}, \texttt{fig2\_spectrum\_phase\_tree.py}, \texttt{fig34\_enrich\_tongue.py}, \texttt{fig5\_ensemble.py}.

\section*{Referenes}
\bibliography{field_closure}

\end{document}